\documentclass[runningheads]{llncs}
\usepackage{graphicx}
\usepackage{url}
\usepackage[ruled,vlined,linesnumbered]{algorithm2e}
\usepackage{color}
\usepackage{multirow}                     

\usepackage{adjustbox}
\usepackage{tabu}
\usepackage[font=small,skip=0pt]{caption}

\begin{document}
\title{Deep Compressed Pneumonia Detection for Low-Power Embedded Devices}

\author{Hongjia Li\inst{1} \and
Sheng Lin\inst{1} \and Ning Liu\inst{1} \and Caiwen Ding\inst{2}
\and Yanzhi Wang\inst{1}\vspace{-1ex}}

\institute{Northeastern University, Boston MA 02115, USA \\
\email{\{li.hongjia, lin.sheng, liu.ning\}@husky.neu.edu, yanz.wang@northeastern.edu}
\and
University of Connecticut, Storrs CT 06269, USA \\
\email{caiwen.ding@uconn.edu} }
\maketitle              
\begin{abstract}

Deep neural networks (DNNs) have been expanded into medical fields and triggered the revolution of some medical applications by extracting complex features and achieving high accuracy and performance, etc. On the contrast, the large-scale network brings high requirements of both memory storage and computation resource, especially for portable medical devices and other embedded systems. In this work, we first train a DNN for pneumonia detection using the dataset provided by RSNA Pneumonia Detection Challenge \cite{rsna_dataset}. To overcome hardware limitation for implementing large-scale networks, we develop a systematic structured weight pruning method with filter sparsity, column sparsity and combined sparsity. Experiments show that we can achieve up to 36x compression ratio compared to the original model with 106 layers, while maintaining no accuracy degradation. We evaluate the proposed methods on an embedded low-power device, Jetson TX2, and achieve low power usage and high energy efficiency.

\end{abstract}
\section{Introduction}

There are approximately 450 million people globally (about 7{\%} of the population in the world) suffering from pneumonia, and results in about 4 million deaths per year \cite{ruuskanen2011viral,lodha2013antibiotics}. In the United States, pneumonia accounts for over 500,000 visits to emergency departments \cite{online1} and over 50,000 deaths in 2015 \cite{online2}, keeping the ailment on the list of top 10 causes of death in the country. To accurately diagnose and localize pneumonia, a general diagnostic process requires review of a chest radiograph (CXR) by highly trained specialists and confirmation through clinical history, blood exams and vital symptoms.

To improve the efficiency and reach of diagnostic services, many researchers have extensively studied from medical fields and also computer aided design. In the past years, DNNs have been experiencing a rapid and tremendous progress thanks to the new era of big data. Especially for computer vision problems, deep learning and large-scale annotated image datasets drastically improved the performances of object recognition, detection and segmentation. Through the training processing based on large-scale datasets, DNNs can rapidly learn the complex features and provide helpful functions of diagnose and localization. Many recent works have discussed medical image detection using large-scale  neural  networks. Based on Chest X-ray dataset \cite{wang2017chestx}, recurrent neural cascade model proposed by \cite{shin2016learning}, CheXNet developed by \cite{rajpurkar2017chexnet}, and Text-Image Embedding network (TieNet) introduced by \cite{wang2018tienet}. Despite the promising results obtained by these works, one of the biggest challenges is that all these networks adopted a deep architecture with multiple layers, leading to a large memory storage and computation resource requirement. These make it difficult to implement large DNN models in portable medical devices and embedded systems \cite{lin2018fft,li2019admm}. 


In order to deploy DNNs on these embedded devices, DNN model compression techniques such as weight pruning, have been proposed for storage reduction and computation acceleration. 
Recently, works such as \cite{han2015deep,zhang2018systematic} have made breakthrough on the weight pruning methods for DNNs while maintaining the network accuracy. 
However, the network structure and weight storage after pruning become highly irregular and therefore the storage of indexing is non-negligible, which undermines the compression ratio and the performance. Therefore, the structured pruning is proposed to incorporate structured sparsity into the weight pruning algorithm \cite{he2017channel,wen2016learning}. The structured sparsity of DNN introduced by pruning methods is hardware-friendly, and it efficiently improves the evaluation of DNNs on embedded devices.

\vspace*{-5pt}
\begin{figure}[t]
	\centering
	\includegraphics[width = 0.95\columnwidth]{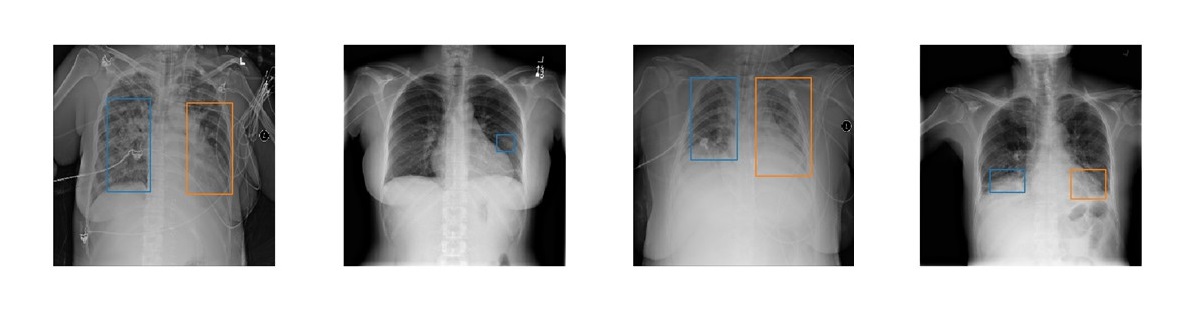}
	\setlength{\belowcaptionskip}{-10pt}
	\caption{Examples of pre-processed data. The boxes showed in figure denotes the detected pneumonia. }
	
	\label{fig:dataset}
\end{figure}  

In this work, we develop a pneumonia detector based on you only look once (YOLO) \cite{redmon2016you}. We select a dataset provided by RSNA Pneumonia Detection Challenge \cite{rsna_dataset}. In the pre-processing stage, the labeled images are resized to $320 \times 320$, along with the corresponding coordinates of bounding boxes, as shown in Figure \ref{fig:dataset}. YOLOv3 \cite{redmon2018yolov3} is adopted as the base feature detector with our costumed anchor box priors, due to the speed boost and high average precision. It can achieve detection accuracy of 71.23 mAP. Moreover, in order to enhance the network performance, we utilize training optimizations including learning rate warmup, cosine learning rate decay and mixup training. To further maintain the precision obtained by the 106-layer network, we apply the ADMM-based unified model pruning algorithm on the original model, incorporated with structured sparsity (filter-wise sparsity and column-wise sparsity). Experimental result shows that without accuracy loss, our YOLOv3-based network can be pruned up to 36x. The number of parameters is reduced from 61.5M to 1.7 M, which undoubtedly reduces the memory storage and computation resource requirement for embedded systems. To validate our proposed method, we implement our model on Jetson TX2 \cite{tx2}, and it achieves low power usage and high energy efficiency. Therefore, it verifies that our proposed method is very suitable for pneumonia detection with the characteristics of real-time and low-power on portable medical devices.

\section{Model Design}

\subsection{YOLOv3}
YOLO is an unified, real-time object detection framework. Compared with other object detection classifiers, YOLO frames object detection as a regression problem to spatially separated bounding boxes and associated class probabilities \cite{redmon2016you}. Recently, two improved versions of YOLO have been developed, namely YOLO9000 \cite{redmon2017yolo9000} and YOLOv3 \cite{redmon2018yolov3}. In this work, we adopt YOLOv3 based detector due to its speed boost and high average precision. 

YOLOv3 is a fully convolutional network, containing 75 convolutional layers, with skip connections and upsampling layers. The YOLOv3 adopts a convolutional layer with stride 2 as downsampling layer instead of pooling layer. A custom deep architecture Darknet-53 is utilized as the feature extractor since it can achieve a promising performance while with fewer floating point operations and more speedup \cite{redmon2018yolov3}. In our work, we initialize the weights using a pretrained DarkNet-53 weights based on ImageNet. 

YOLOv3 predicts boxes at 3 different scales. For each scale, detection layers that comprised of convolutional layers are constructed, respectively. The last layer predicts a 3D tensor containing bounding box coordinates, object prediction, and class predictions. In our work, the class number is 1 and the number of predicted boxes at each scale is 3, thus the tensor is $N \times N \times \left [ 3 * \left ( 4 + 1 + 1 \right ) \right ] $ for the 4 bounding box offsets, 1 object prediction, and 1 class predictions. YOLOv3 predicts bounding boxes using dimension clusters as anchor boxes. The network predicts 4 coordinates for each bounding box. K-means clustering is adopted to determine our anchor boxes. Same as YOLOv3, we choose 9 clusters and 3 scales. On our data, we modify the 9 clusters as following: $(40\times39), (63\times49), (48\times69), (75\times74), (58\times102), (83\times108), (67\times148), (89\times154), (94\times202)$.

\subsection{Training optimization}
Inspired by \cite{xie2018bag}, we absorb several training optimization methods to enhance the network performance. 
\textbf{Learning rate warmup:} Instead of using a too large learning rate directly at the beginning, we use a small learning rate and then smooth back to the initial learning rate. To be specific, we use a gradual warmup strategy, which increases the learning rate from 0 to the original initial learning rate linearly.
\textbf{Cosine learning rate decay:} For the learning rate decay, a cosine annealing strategy is applied, in which the learning rate gets decreased from the initial value to 0 by the following function: $lr_{t} = 0.5 * (1+ cos(t\pi / T))lr_0$, where $t$ denotes the current batch and $T$ denotes the total number of batches, and $lr_0$ is the initial learning rate. 
\textbf{Mixup:} For data augmentation, we adopt mixup method, in which each time we randomly sample two examples $(x_i, y_i)$ and $(x_j. y_j)$. Then a new example is obtained by a weighted linear interpolation of these two examples: $x' = \lambda x_i+(1-\lambda)x_j$, $y' = \lambda y_i+(1-\lambda)y_j$, where $\lambda \in [0,1]$ is a random number drawn from the $Beta(\alpha, \alpha )$ distribution. The new example $(x', y')$ will be used as our training data.

\section{Model Compression}

\subsection{Unified weight pruning algorithm}
We develop an unified systematic framework containing three phases: pre-pruning, masked mapping and retraining. The objective of the weight pruning is to minimize the loss function while satisfying the weight constraints, the whole problem is defined as:  

\begin{equation}\label{eqn:goal}
\centering
{minimize\ } f_{Loss} \big( \{{{W}}_{i}\}_{i=1}^N, \{{{b}}_{i} \}_{i=1}^N \big),\;  subject \; to\; 
{{W}}_{i}\in {\mathcal S }_{i},\  i = 1, \ldots, N.
\end{equation}

where ${W}_{i}$ and ${b}_{i}$ denotes the sets of weights and biases of the $i$-th (CONV or FC) layer in an $N$-layer DNN, respectively. The set ${\mathcal{S}}_{i}=\big\{{{W}_{i}\big|card({{W}}_{i})\le \alpha_i\big\}}$ denotes the constraint for weight pruning, and `card' refers to cardinality. It meets the goal that the number of non-zero elements in ${{W}}_{i}$ is limited by $\alpha_i$ in layer $i$.

In the pre-pruning phase, we add the ADMM-based regularization on an original DNN model. The regularization is operated by introducing auxiliary variables ${{Z}}_{i}$'s, and dual variables ${{U}}_{i}$'s. In each iteration, while keeping on minimizing network regularized loss, we also reduce the error of Euclidean projection from  ${{W}}_{i}^{k+1}+{{U}}_{i}^{k}$ onto the set ${\mathcal{S}}_{i}$. Because under the constraint that $\alpha_i$ is the desired number of weights after pruning in the $i$-th layer, the Euclidean projection can keep $\alpha_i$ elements in ${{W}}_{i}^{k+1}+{{U}}_{i}^{k}$ with the largest magnitudes and set the remaining weights to zeros. Then the dual variables ${{U}}_{i}$ is updated as following: ${{U}}^{k+1}_{i} = {{U}}^{k}_{i} + {{W}}^{k+1}_{i} - {{Z}}^{k+1}_{i}$. In the second phase, with the obtained intermediate ${{W}}_{i}$ solutions, we first perform the Euclidean projection (mapping) to satisfy that at most $\alpha_i$ weights in each layer are non-zero. And then in the retraining phase, the zero weights are gradient masked and non-zero weights are retrained using training sets to restore partial accuracy.

\vspace*{-5pt}
\begin{figure}[t]
	\centering
	\includegraphics[width = 0.95\columnwidth]{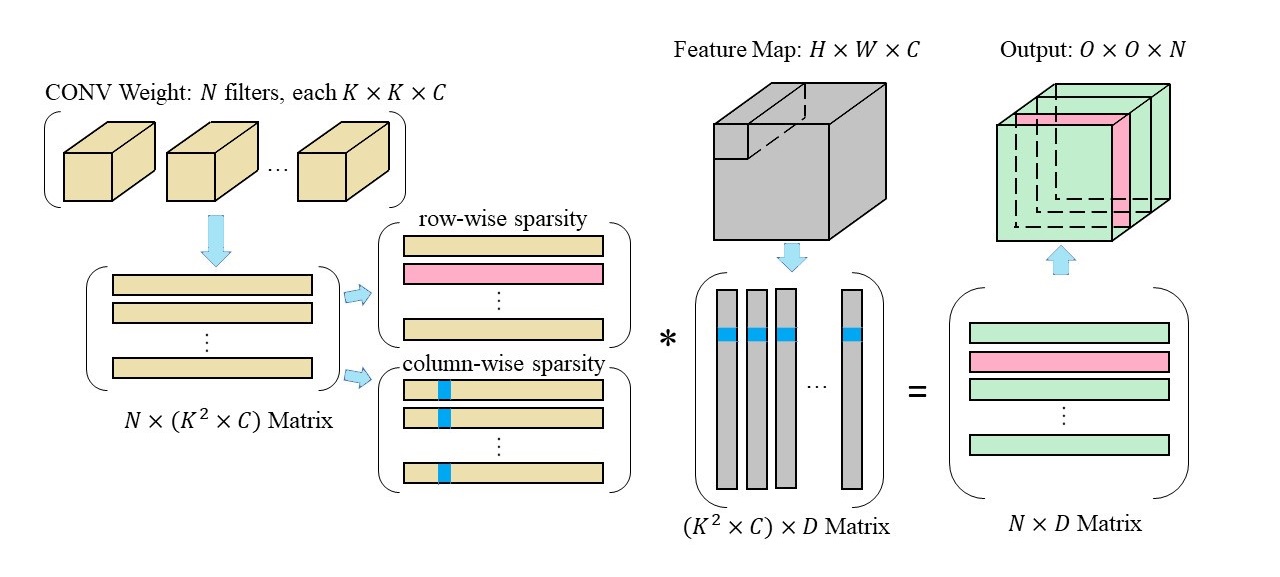}
	\setlength{\belowcaptionskip}{-10pt}
	\caption{Examples of GEMM in CONV layer and effect of structured sparsities.}
	\label{fig:gemm}
\end{figure}
\vspace*{-5pt}

\subsection{Structured pruning}
As mentioned before, irregular pruning methods introduce extra storage for index and undermines the  compression  ratio  and  the  performance. In order to develop an algorithm more friendly on hardware implementation, we incorporate structured pruning with the unified weight pruning algorithm. 

In a typical convolutional layer, there are two structured sparsities: filter-wise sparsity, channel-wise sparsity, and shape-wise sparsity. For fully-connected layers, there are two types: row-wise sparsity and column-wise sparsity. We mainly focus on compressing convolutional layer in our design, since it is the most computationally intensive layer in current DNNs and our model is a fully convolutional network.

During the convolutional computation, the feature map tensor and weights tensor are converted to 2D matrices and performed the general matrix multiplication (GEMM), as shown in Figure \ref{fig:gemm}. Filter-wise sparsity corresponds to row-wise sparsity, while channel-wise sparsity and shape-wise sparsity correspond to column-wise sparsity. Therefore, filter pruning leads to reducing the number of rows of matrix, and correspondingly, channel and shape pruning result in the reduction of column number. The process of our structured pruning method can be explained as follows.

\subsubsection{Filter pruning}

\hspace{0.5cm} As we mentioned in Equation \ref{eqn:goal}, the constraint set ${\mathcal{S}}_{i}$ here indicates the number of nonzero filters in ${{W}}_{i}$ that is less than a predefined value $\alpha_i$. To determine the limited number of nonzero filters, we perform $l2 \ norm$ on each filter and select the $\alpha_i$ filters with most magnitude and set the remaining as zero. 

\subsubsection{Column pruning}

\hspace{0.5cm} In the pre-pruning phase, we prune the convolutional weight by first converting 4D weight tensor into a 2D matrix. Therefore,  the constraint set ${\mathcal{S}}_{i}$ for column pruning indicates the number of nonzero column in converted ${{W}}_{i}$ that is less than a threshold value. The largest $\alpha_i$ columns evaluated by $l2 \ norm$ are kept and the remaining column values are set to zero. 

\subsubsection{Combined pruning}

\hspace{0.5cm} To take advantage of utilization in structured pruning on hardware implementation, we propose a approach by combination of these two structured pruning, which decreases the dimension in GEMM while still maintaining a full matrix. We first perform either one type pruning, filter for example. With the filter-pruned model, we first mask the zero filters and then perform the column pruning. In this way, we can keep the desired number of nonzero filter and obtain a higher sparsity on the column-wise.

\section{Experimental Results}
In this section, we evaluate the proposed model compression technique, starting from original model training, systematic structured weight pruning, and the hardware implementation on embedded device. 

\subsection{Data preprocessing}
In our project, we use the dataset provided by RSNA Pneumonia Detection Challenge \cite{rsna_dataset}. The dataset is derived from National Institutes of Health Clinical Center for publicly providing the Chest X-Ray dataset \cite{wang2017chestx}. In our experiments, only the labeled images are selected, loaded from Digital Imaging and Communications in Medicine (DICOM) image format and resized into $320 \times 320$ from the original $1024 \times 1024$. The corresponding coordinates are also re-calculated from the original size. The whole dataset contains 6,002 images, of which 5,400 are considered as our training dataset and the remaining 602 are test dataset.

\subsection{Model training}
We apply the weight pruning method and train the pneumonia detector on Nvidia GeForce GTX2080 using Pytorch. During the  training, we warmup our learning rate from $10^{-5}$ to our initial learning rate $10^{-3}$ during the first epoch. In the rest epochs, the learning rate decreased from $10^{-3}$ to $4^{-8}$ using the cosine function. The $\alpha$ for the $Beta$ distribution in data mixup is 0.2.

\subsection{Model evaluation}
To evaluate the performance of the model, we use mean average precision (mAP) at different intersection over union (IoU) thresholds. The metric sweeps over a range of IoU thresholds, at each point calculating an average precision value. The threshold values range from 0.4 to 0.75 with a step size of 0.05: (0.4, 0.45, 0.5, 0.55, 0.6, 0.65, 0.7, 0.75). To be specific, if we use 0.5 as the threshold, only when IoU is greater than 0.5 the object can be considered as detected. The result of original model is listed on the first row in Table \ref{result} under different IoU thresholds. When IoU threshold is 0.5, we can achieve detection accuracy of 71.23 mAP.

Next, the unified structured weight pruning method is applied on filter pruning, column pruning and combined pruning, respectively. The detailed evaluation results of models with various prune ratio are shown in Table \ref{result}. Without accuracy loss, the prune ratio can be increased up to 36x. For this model, we prune 3.56x filters and 9.68x columns. The size of model parameters is reduced from 61.5 M to 1.7 M, which results the model storage saved from 246.4 MB to 6.84 MB. The original floating point operations (FLOPs) is 38.63 Bn. In total, the FLOPs can be significantly reduced to 1.32 Bn. In this way, not only the requirement of memory storage and computation resource decreased, but also facilitate acceleration on embedded devices. 

\begin{table}[]
\centering
\footnotesize
\caption{Localization accuracy (mAP) using IoU where T(IoU)={0.4, 0.45, 0.5, 0.55, 0.6, 0.65, 0.7, 0.75}.}
\label{result}
{
\begin{tabular}{c|c| p{0.9cm}p{0.9cm}p{0.9cm}p{0.9cm}p{0.9cm}p{0.9cm}p{0.9cm}p{0.9cm}}
\hline
\hline
\multicolumn{2}{c|}{T(IoU)}                       & 0.4 & 0.45 & \bf{0.5} & 0.55 & 0.6 & 0.65 & 0.7 & 0.75 \\
\hline \hline
\multicolumn{2}{c|}{Original model}               & 81.2  &  76.3    &  \bf{71.2}   &   63.4   &  54.7   &   42.3   &   30.7  &   19.1   \\
\hline
\multirow{2}{*}{Filter pruned}         
                                 & 11.55x        &  81.4  &  75.9    &   \bf{71.5}  &  63.8    &  53.0   &  40.9      &  29.7   &  17.9    \\
                                & 16.26x        &  80.6  &  76.2    &   \bf{71.2 } &   62.5   &  53.6   &  42.3    &  30.0   &  18.7    \\
                                & 19.33x        &  80.7  &  76.1   &   \bf{71.1}  &   62.3   &  52.9   &   41.2   &  30.0   &  18.6    \\
\hline
\multirow{2}{*}{Column pruned} 
                                & 11.60x        &  81.2  &  76.9   &  \bf{71.9}   &   64.5   &  53.3   &   41.8   &  29.8   & 18.4    \\
                                & 16.36x      & 80.7   &  76.0    &   \bf{71.3} &  64.1    &  55.9   &  41.5 &  28.4   &  19.1   \\
                                 & 19.55x      &  80.6  &  76.1  & \bf{71.0}  &  63.7    &  53.5   &  42.3    &  29.7   &  18.7   \\
\hline
\multirow{2}{*}{Combined pruned} 
                                &  \bf{36.02x}      &  \bf{81.2}   &   \bf{76.3}   &  \bf{71.0}   &   \bf{63.5}   &  \bf{53.4}   &   \bf{42.8}   &   \bf{31.2}  &  \bf{19.3}   \\
                                &  51.97x      &  81.0  &   76.0   &  \bf{70.6}   &    62.8  &  53.0   &  41.7    &  29.0   &  18.3   \\
\hline
\end{tabular}
}
\end{table}

\subsection{Hardware implementation}

To validate our method on the embedded low-power devices, we implement our pruned model on Jetson TX2, which is considered as the fastest, most power-efficient embedded AI computing device \cite{tx2}. It's built by a 256-core NVIDIA Pascal-family GPU and the memory is 8 GB with 59.7 GB/s bandwidth. The power consumption of our model is 7.3 W and the energy efficiency is 0.69 IPS/W. The low power usage and high energy efficiency show a high feasibility and compatibility of our weight pruning method on DNN for low-power real-world devices. 

\section{Conclusion}

In this work, we developed a YOLOv3-based detector for pneumonia detection with 71.23 mAP. In order to reduce the storage memory and computational resource requirement by the 106-layer fully convolution network, we applied a systematic structured weight pruning method on filter sparsity, column sparsity and combined sparsity. Without accuracy loss, the prune ratio can achieve up to 36x, which reduce the model size from 
61.5 M to 1.7 M. To validate our method on the real-world low-power device, we implemented and evaluated our model on Jetson TX2, which resulted a low power usage and high energy efficiency.

\bibliographystyle{splncs04}
\bibliography{references}

\end{document}